# SIMULATION TO TRACK 3D LOCATION IN GSM THROUGH NS2 AND REAL LIFE

Anand Gupta, Harsh Bedi , MS Don Bosco, Vinay Shashidhar

Department of Computer Engineering
Netaji Subhas Institute of Technology,
Delhi University, New Delhi, India

anand@coe.nsit.ac.in, harshbedi@gmail.com, msdonbosco@gmail.com,
vinayshashidhar@gmail.com

## ABSTRACT

*In recent times the cost of mobile communication has dropped significantly leading to a dramatic increase in mobile phone usage. The widespread usage has led mobiles to emerge as a strong alternative for other applications one of which is tracking. This has enabled law-enforcing agencies to detect over-speeding vehicles and organizations to keep track its employees. The 3 major ways of tracking being employed presently are (a) via GPS [1] (b) signal attenuation property of a packet [3] and (c) using GSM Network [2]. The initial cost of GPS is very high resulting in low usage whereas (b) needs a very high precision measuring device. The paper presents a GSM-based tracking technique which eliminates the above mentioned overheads, implements it in NS2 and shows the limitations of the real life simulation. An accuracy of 97% was achieved during NS2 simulation which is comparable to the above mentioned alternate methods of tracking.*

## KEYWORDS

*Global System for Mobile Communication (GSM), Network, Global Positioning System (GPS), Signal Attenuation, Triangularization, Network Simulator, Signal Mast.*

## 1. INTRODUCTION

In modern world, location estimation has almost become a necessity. It helps law enforcing agencies to keep track of over speeding vehicles [6], and assists organizations in keeping a check on inefficient employees [8].

Location estimation has almost become a necessity in modern times. Law enforcing agencies keeping track of over speeding vehicles [6] and organizations keeping a check on inefficient employees [8] are just some examples where this method has proved to be more than just useful.

GPS, the classical way of tracking, involves determining the latitude and longitude position of an object with a very high accuracy. However this requires a dedicated device for communicating with the GPS access points. Thus it increases the cost to make avail of this service.

Hence in the recent years the focus has now shifted on finding other ways of tracking**.** This has led to the emergence of alternate technologies. GSM Network technique is one of them. Use of GSM network for tracking, in the initial days, did not prove useful because of the lack of widespread use of mobiles. But of late the number of mobile users has increased dramatically due to the decrease in cost of mobile communication, as a result of which mobile phones have become a potent tool for tracking.

There are two approaches to make use of mobiles as a tracking device. One of the approaches makes use of signal attenuation (energy dissipation) [4] property of a packet. This method makes use of the fact that energy, of the packet, dissipated is related to the distance travelled by it. However this method requires a highly precise and accurate energy measuring device. This again increases the cost of the tracking system.

Another alternate technique uses the GSM network. It makes use of the signal masts to determine the position of the mobile. The GSM network technology works by broadcasting a request packet to all the towers. The towers on receiving the request packet sends a acknowledge packet back to the mobile. Tracking the mobile requires use of the time-interval for the round trip of the packet. Greater the precision of the time-interval, greater is the accuracy of the tracking. As no extra devices are required for this process it eliminates the cost overheads required by the other two methods

The organization of the remaining paper starts with the system design and the algorithm in which we have described how tracking is performed using the GSM network technique. The calculation and the results which come from the NS2 implementation precede the error calculation. The paper then covers the real life simulation of the GSM network technique and reports the difficulties faced during the experiment. The future work section covers our aspirations, and the direction in which we would like to take the research work. We end the paper with our findings on the topic and the measures which should be taken to make this paper feasible for mobile tracking applications.

## 2. MOTIVATION

The existing tracking techniques suffer from high cost overheads whereas the GSM tracking technique is quite cost-efficient. This has motivated us to pursue the latter tracking technique which if implemented on a large scale it could solve many real life problems such as

- Thefts of expensive handsets are on the rise nowadays which results in the loss of important contacts. An algorithm which can track the mobile without any extra cost to the pocket is the need of the hour.

- This approach would be helpful in firms wherein it is very important for the seniors who want to keep a strict vigil on their juniors [8].

- An algorithm which could determine the speed of a vehicle without installing expensive devices would help the law enforcing agencies [7].

The above-mentioned applications have motivated us to contribute to the GSM-based mobile tracking which is as follows.

## 3. CONTRIBUTION

Our contribution towards the GSM tracking algorithm are as follows

a. This paper is an extension to our previous paper titled 'Accuracy of 3D location computation in GSM through NS2' in which we had only simulated the algorithm on NS2. No real-life analysis was performed.

b. In this paper we are proposing an algorithm that drastically reduces the cost involved in tracking and locating the position of a mobile user. Even though privacy issue could be raised the algorithm is self sufficient to tackle the issue

c. The algorithm proposed will be a very important tool in the hands of the law enforcing agencies as they can now easily get to know the whereabouts of a speeding vehicle and the exact speed at which it is travelling. If could discreetly this would also help companies keep track of their employees.

d. This algorithm drastically reduces the cost involved in tracking and locating the position of a mobile user. Even though privacy issue could be raised the algorithm is self sufficient to tackle the issue.

To continue with the tracking technique we proceeded to simulation of the network in NS2.

## 4. NS2 SIMULATION

NS2 is an open-source discrete event network simulator. It supports an array of popular network protocols, offering simulation results for wired and wireless networks alike. It can be also used as limited-functionality network emulator. Hence, NS2 provides a good platform for simulating the GSM network and testing our proposed algorithm.

Our algorithm specifies the following protocol.

### 4.1. Protocols for Location of GSM Mobile User

The workflow diagram for the protocol employed for the simulated algorithm is depicted in figure 1. The Steps followed are mentioned below.

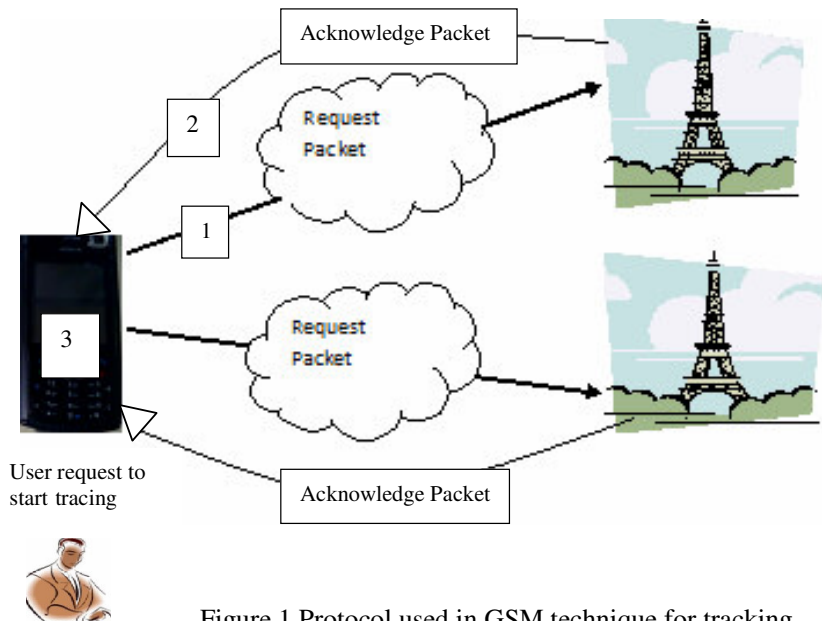

Figure 1 Protocol used in GSM technique for tracking

1. Tower sends a broadcast signal(with a timestamp field that contains the exact time of transmission of packet) directions so that nearby towers may accept this signal

2. Towers accept this packet and sends an acknowledge packet which contains the co-ordinates of the tower as well as the original timestamp and uni-cast it back to the mobile

3. The mobile receives the acknowledge packet and calculates the time interval for the round trip.

## 4.2. Assumption

The protocol mentioned in section 4.1 was implemented with the following assumptions

a) The time taken for sending and receiving the signal from the mobile to the tower is assumed to be so small that user remains stationary in that time frame.

b) The mobile is at any given point of time able to successfully communicate with all the nearby towers.

c) There is no interference with the signals and the signal from the nearest one reaches the first and so on.

d) Signal strength of packets is uniform in all directions into which it is directed.

e) During the journey there is no such place which does not come under the region of towers.

f) When 2 or more packets are reaching a tower or a mobile, at any time instant, the mobile or tower can write the timestamps on the packets together at the same time

## 4.3. Algorithm

With the above stated assumptions we started the implementation of the algorithm. After receiving a trigger from the user, **Algorithm1** Track_init() initializes the tracking process. It broadcasts request packets, containing the time of transmission to all the towers in its range .On receiving the request packet, **Algorithm 2** Tower_response() method acknowledges by adding its co-ordinates to the packet header along with the time of receiving of the request packet. This uni-cast communication is done between the tower and the mobile from where the request packet originated. Finally, **Algorithm 3** Mobile_coord() method calculates the position of mobile by using the response packets from first three towers.

---

**Algorithm** 1: Track_init()

---

    **Input**    **:** Tracking request signal.
    **Output**  **:** Broadcasting request packets to the towers.
    **Method**   **:** Mobile broadcast the request packets.

---

    1.  while (track_mobile() == NO)
    2.      ;   // wait
    3.  for I = 1 to N
    4.  request[i] = new packet()
    5.  request[i].timestamp = current_time()
    6.  request[i].mob_id = mobile_id
    7.  send_tower_request_packet(request[i])
    8.  end

---

**Algorithm2 :** Tower_response ( )

---

    **Input**    **:** Request packet from mobile.

**Output :** Acknowledge packet from tower containing its coordinate.
**Method :** Tower responds with a reply packet adding its coordinate.

---

1. while (received_packet() == NO)
2. ;    // wait
3. mobile_id = get_mobile_id_read_packet()
4. old_time_stamp = get_time_stamp_read_packet()
5. acknowledge = new packet()
6. acknowledge.destination = mobile_id
7. acknowledge.tower = tower_id
8. acknowledge.tower_coord = tower_co-ordinates
9. acknowledge.timestamp = old_time_stamp
10. send_mobile_acknowledge_packet(acknowledge)
11. end

---

**Algorithm 3:** Mobile_coord()

---

**Input     :** Acknowledge packet from towers.
**Output:** Coordinates of the Mobile.
**Method: Evaluates** the coordinates of the mobile.

---

1. for I=1 to 3
2. while (receive_ack_packet() == NO)
3. ;    // wait
4. time_prop = current_time() – acknowledge.timestamp
5. distance[i] = ((time_prop-delays)*speed_of_light)/2
6. coord[i] = acknowledge.tower_coord
7. calculate_cord()

The following experimental results were obtained on simulating the GSM algorithm.

## 4.4. Experimental results

The GSM tracking technique is implemented in network simulator (NS2) in accordance with the protocols mentioned above. The results were collected from the   simulation and the calculations have been shown in the subsequent sections.

### 4.4.1. Topography used

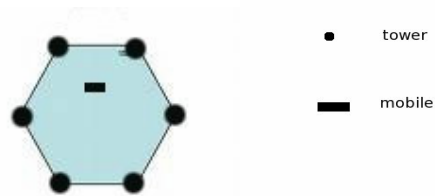

Figure 2 Depicts a typical mobile cell

To enable the GSM based handset the cellular service providers generally divide the region into small regions called cells, which are generally hexagonal in shape. A typical cellular cell has been shown in figure 2.

The tower numbering starts from 0 and goes till 5 with the leftmost node being 0 and the rest numbered in clockwise fashion as shown in figure 2.

TOWER 0: (0, 5, 0)              TOWER 1: (0, 15, 0)
TOWER 2: (8.66, 0, 0)          TOWER 3: (8.66, 20, 0)
TOWER 4: (17.32, 5,0)          TOWER 5: (17.32, 15, 0)
                Mobile Position: (4, 3.5, 0)

Figure 3 NS2 simulation results

## 4.7. Formula Used

Time interval between transmission of packet = a constant delay + time required for and it's receiving by tower propagation

Time required for propagation = distance travelled / speed of light

The table given below has been tabulated using the results from the NS2 simulation (figure 3)

| Tower | Time interval | Distance calculated | Actual distance | % error |
|---|---|---|---|---|
| 0 | 7.58014*10-4 | 4.2 | 4.272 | 2.20 |
| 2 | 7.58019*10-4 | 5.7 | 5.828 | 1.02 |
| 1 | 7.58041*10-4 | 12.3 | 12.176 | -1.68 |
| 4 | 7.58044*10-4 | 13.2 | 13.404 | 1.52 |
| 3 | 7.58056*10-4 | 16.8 | 17.145 | 2.01 |
| 5 | 7.58058*10-4 | 17.4 | 17.596 | 1.11 |

Table 1 Tabulated Result

The calculations which have been performed are shown below

## 4.8. Calculation result

For Calculation purposes the following equations were used

$T1(x1, y1, z1)$, $T2(x2, y2, z2)$, $T3(x3, y3, z3)$ are the coordinates of the closest three towers

$M(x,y,z)$ are the coordinates of the mobile phone

$$(x-x1)^2 + (y-y1)^2 + (z-z1)^2 = r1^2 \quad \ldots\ldots\ldots\ldots \text{(1)}$$

$$(x-x2)^2 + (y-y2)^2 + (z-z2)^2 = r2^2 \quad \ldots\ldots\ldots\ldots \text{(2)}$$

$$(x-x3)^2 + (y-y3)^2 + (z-z3)^2 = r3^2 \quad \ldots\ldots\ldots\ldots \text{(3)}$$

Subtracting Eq: (1) –Eq: (2) we get

$$\lambda_1 x + \mu_1 y + \mu_1 z = \xi_1 \quad \ldots\ldots\ldots\ldots \text{(4)}$$

Subtracting Eq: (2) –Eq: (3) we get

$$\lambda_2 x + \mu_2 y + \sigma_2 z = \xi_2 \quad \ldots\ldots\ldots\ldots \text{(5)}$$

Subtracting Eq: (3) –Eq: (1) we get

$$\lambda_3 x + \mu_3 y + \sigma_3 z = \xi_3 \quad \ldots\ldots\ldots\ldots \text{(6)}$$

Where $\lambda_i, \mu_i, \mu_i, \xi_i$ are scalars obtained on subtraction of the equations

$$\begin{bmatrix} \lambda 1 & \mu 1 & \sigma 1 \\ \lambda 2 & \mu 2 & \sigma 2 \\ \lambda 3 & \mu 3 & \sigma 3 \end{bmatrix} \begin{bmatrix} x \\ y \\ z \end{bmatrix} = \begin{bmatrix} \xi 1 \\ \xi 2 \\ \xi 3 \end{bmatrix}$$

Hence,

$$\begin{bmatrix} x \\ y \\ z \end{bmatrix} = \begin{bmatrix} \lambda 1 & \mu 1 & \sigma 1 \\ \lambda 2 & \mu 2 & \sigma 2 \\ \lambda 3 & \mu 3 & \sigma 3 \end{bmatrix}^{-1} \begin{bmatrix} \beta 1 \\ \beta 2 \\ \beta 3 \end{bmatrix}$$

Smallest 3 distance/smallest 3 time delays:

1. 4.2
2. 5.7
3. 12.3

Value found is

       X co-ordinate=3.84
       Y co-ordinate =3.31
       Z co-ordinate =0

**Error calculation:**

       In x- co-ordinate: 3.79%
       In y co-ordinate: 5.21%
       In z co-ordinate: 0%

## 5. REAL LIFE SIMULATION

Having completed the NS2 simulation, we proceeded to model the real-life scenario using 4 laptops. The workflow diagram of the experiment has been shown in Figure 4.

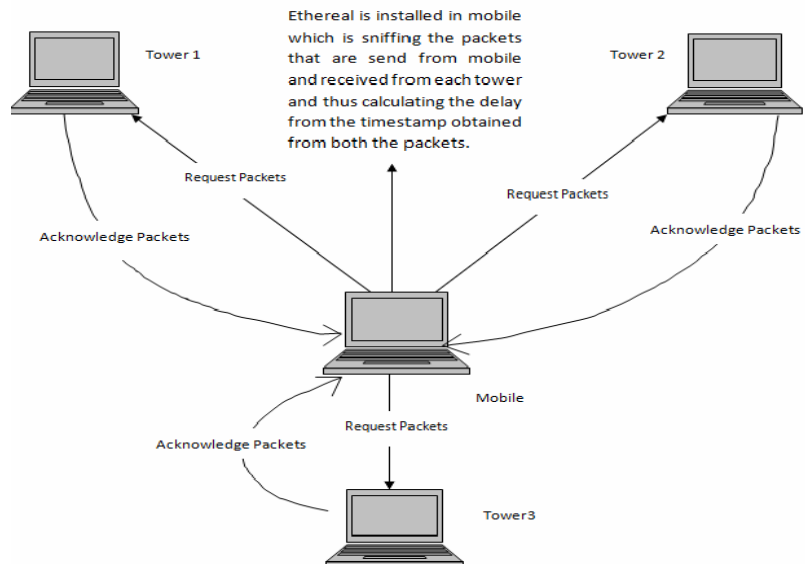

Figure 4 Workflow diagram of the real-life modelling of GSM network using 4 laptops

The steps followed are:

1. Simulation was carried out by setting up the wireless ad-hoc network between the 4 laptops treating 3 as towers and 4[th] one as the mobile.

2. A GSM like network was simulated by exchanging the packets between the laptops using ICMP protocol.

3. The packets that were exchanged between the laptops acting as towers and mobile were sniffed from the OSI physical layer using ethereal packet sniffer.

4. The timestamp obtained from the sniffed packets were used to calculate the total network delay involved in sending and receiving of ICMP packet between the towers and the mobile.

5. The delay due to the kernel (i.e. the delays involved in the kernel to move from the application OSI layer to the physical OSI layer) can be determined by self-pinging(ping 127.0.0.1) and determine the time-delays involved.

6. The kernel delays are subtracted from the total network delay to get the propagation delay which is the time interval involved in the propagation of the packet with the speed of light.

The observations hence obtained were compared with those from NS2.

## 5.1. Observations obtained from Ethereal:

The results on pinging from mobile to tower1 have been recorded in figure 5

Figure 5. Ping from mobile to Tower1

The results on pinging from mobile to tower2 have been recorded in figure 6

Figure 6. Ping from mobile to Tower2

The results on pinging from mobile to tower3 have been recorded in figure 7

| | | | | |
|---|---|---|---|---|
| 21 9.000041 | 169.254.118.52 | 169.254.65.4 | ICMP | Echo (ping) request |
| 22 9.000015 | 169.254.65.4 | 169.254.118.52 | ICMP | Echo (ping) reply |
| 23 10.000046 | 169.254.118.52 | 169.254.65.4 | ICMP | Echo (ping) request |
| 24 10.000740 | 169.254.65.4 | 169.254.118.52 | ICMP | Echo (ping) reply |
| 25 11.000089 | 169.254.118.52 | 169.254.65.4 | ICMP | Echo (ping) request |
| 26 11.000802 | 169.254.65.4 | 169.254.118.52 | ICMP | Echo (ping) reply |
| 27 12.000055 | 169.254.118.52 | 169.254.65.4 | ICMP | Echo (ping) request |
| 28 12.000710 | 169.254.65.4 | 169.254.118.52 | ICMP | Echo (ping) reply |
| 29 13.000059 | 169.254.118.52 | 169.254.65.4 | ICMP | Echo (ping) request |
| 30 13.000731 | 169.254.65.4 | 169.254.118.52 | ICMP | Echo (ping) reply |
| 31 14.000065 | 169.254.118.52 | 169.254.65.4 | ICMP | Echo (ping) request |
| 32 14.000843 | 169.254.65.4 | 169.254.118.52 | ICMP | Echo (ping) reply |
| 33 15.000116 | 169.254.118.52 | 169.254.65.4 | ICMP | Echo (ping) request |
| 34 15.000806 | 169.254.65.4 | 169.254.118.52 | ICMP | Echo (ping) reply |

Figure 7. Ping from mobile to Tower3

## 5.2. Formula used

The hypothesis of GSM tracking proposed by us is based upon the following equations which are as follow

$$T = \alpha + \text{time\_prop} \quad .... (7)$$

Where

| | |
|---|---|
| α | be constant internal delays, |
| time_prop | be time involved due to propagation of packet only, |
| T | be the turn-around time between sending of ICMP packet by the mobile and receiving of the reply packet by the same mobile laptop |
| distance[i] | be the distance travelled by ith packet, |
| c | be the speed of light. |

Also,                                    $\text{time\_prop} = D/c.... (8)$

Where

| | |
|---|---|
| D | be the total distance of propagation by the packet wave |
| And     c | be the speed of light |

## 5.3. Calculations

The calculation were carried out in accordance to the formulae given in section 6.2 and the observations given in figure 5,figure 6 and figure 7.

| S.No | Time Interval between sending and receiving of ICMP packet (in milliseconds) |
|---|---|
| 1. | 0.783 |
| 2. | 0.799 |
| 3. | 0.690 |
| 4. | 0.985 |
| 5. | 0.567 |
| 6. | 0.533 |
| 7. | 0.671 |

Table 2. Ping between mobile and tower1

| S.No | Time Interval between sending and receiving of ICMP packet (in milliseconds) |
|------|------------------------------------------------------------------------------|
| 1. | 0.543 |
| 2. | 0.664 |
| 3. | 0.764 |
| 4. | 0.667 |
| 5. | 3.608 |
| 6. | 0.674 |
| 7. | 0.645 |

Table 3. Ping between mobile and tower 2

| S.No | Time Interval between sending and receiving of ICMP packet (in milliseconds) |
|------|------------------------------------------------------------------------------|
| 1. | 0.774 |
| 2. | 0.694 |
| 3. | 0.714 |
| 4. | 0.655 |
| 5. | 0.672 |
| 6. | 0.778 |
| 7. | 0.770 |

Table 4. Ping between mobile and tower 3

## 5.4. Constraints faced

1. The range of wireless ad-hoc network setup using laptops was experimentally found out to be 175 metres.

2. The packet captured by the packet sniffer was allotted a timestamp by the kernel of the operating systems used (Linux kernel version 2.26.x, Windows XP Service Pack and Windows Vista Home Premium) which is only up to precision 6 digits. So, the delay which could be calculated only in micro seconds ($10^{-6}$ seconds).

Maximum Propagation Delay (theoretical)

$$= \frac{\text{Maximum range of wireless ad-hoc network}}{\text{Speed of light}}$$

$$= \frac{(175 \text{ metres})}{(3 \times 10^8 \text{ m/sec})} = 0.5833 \times 10^{-6} \text{ sec.}$$

The above calculation shows that the minimum precision required for the timestamp, to determine the distance between the mobile and the tower using the algorithm specified should be at least in the order of $10^{-7}$ sec.

This time difference cannot be observed by the laptops due to the above mentioned constraints.

## 6. CONCLUSION

This paper implements network based tracking algorithm both in NS2 as well as real-life simulation for tracking GSM mobile devices. To the best of our knowledge no one has simulated this algorithm in NS2. This paper advocates the use of triangularization approach [9] for tracking GSM mobiles. The approach used in simulation comprises of mainly three stages. In first stage, the first three responses of packets from towers were recorded. In second stage, time difference between the sending and receiving of packet was calculated. Finally in the last stage, the mobile location was identified.

The simulation results from NS2 confirm author's intuition (either tracking employees of an organization or detecting over speeding vehicles) that GSM based tracking technique will be best suited for commercial purposes over existing methods.

Further the real-life simulation of the algorithm being limited by the precision of the time-stamp value as provided by the Operating System Kernel of the laptops(which is in the order of micro-seconds) is unable to provide the propagation delays essential for triangularization. However, on being provided with a high-precision timestamp device the author believes that the real-life simulation performed as prescribed in the paper using 4 laptops will behave in accordance with the NS2 simulation results.

## 7. FUTURE WORK

This paper attempts to perform the real life simulation of the GSM network techniques. But due to the constraints mentioned, the simulation could not exactly determine the accuracy of the techniques in real life. We are presently focusing on working our way around the difficulties and trying to come up with a complete analysis that will help us in further understanding the feasibility of the algorithm in real life.

## REFERENCES

[1] R.I. Desourdis, A.K. McDonough, S. C. Merrill, R.M. Bauman, D. A. Neumann, J.A. Lucas, D. Spector, and D. E.Warren, "GPS-based vehicle tracking with meteor-burst telemetry", Proc. IEEE Conf. Military Communications (MILCOM 94), IEEE Press, Oct. 1994, pp.690-694, Vol. 3, doi:10.1109/MILCOM.1994.473883.

[2] Ding-Bing Lin, Rong-Terng Juang, and Hsin-Piao Lin, "Mobile location estimation and tracking for GSM systems", Proc. IEEE Symp. Personal, Indoor and Mobile Radio Communications (PIMRC 04), IEEE Press, Sept. 2004, pp. 2835-2839, Vol. 4, INSPEC Accession Number: 8221818.

[3] H.C. Lee, and S. Junping, "Multilayered Model Strategy for Optimal Mobile Location Tracking", Proc. IEEE Symp. Personal, Indoor and Mobile Radio Communications (PIMRC 97), IEEE Press, Sept. 1997, pp. 1009-1013, Vol. 3, doi:10.1109/PIMRC.1997.627038.

[4] C. Chao-Lin, and F. Kai-Ten, "Hybrid Location Estimation and Tracking System for Mobile Devices",Proc. IEEE Conf. Vehicular Technology Conference (VTC 05), IEEE Press, 2005, pp. 2648-2652, Vol. 4, doi: 10.1109/VETECS.2005.1543815.

[5] Security firm using GPS Sacramento Business Journal - June 24, 2005 by Mark Larson Staff Writer

[6] T. Stamoulakatos, A. Yannopoulos, T. Varvarigou, and E. Sykas, "Hidden Markov Filtering with Microscopic Traffic Modeling for Vehicle Load Estimation in Cellular Networks", In Proceeding of IASTED International Conference on Communication Systems and Networks (CSN 2003), Sept. 1-3, pp. 349-355, 2004.

[7] C. S. Jensen, K. J. Lee, S. Pakalnis, and S. Šaltenis, "Advanced Tracking of Vehicles", In Proc. European Congress and Exhibition on ITS, 12 pages (2005).

[8] T. Sohn, A. Varshavsky, A. LaMarca, M. Y. Chen, T. Choudhury, I. Smith, S. Consolvo, J. Hightower, W. G. Griswold, and E. D. Lara, "Mobility Detection Using Everyday GSM Traces", Proc. LNCS Ubiquitous Computing (Ubicomp 06), LNCS Press, Sept 2006, pp. 212-224, doi: 10.1007/11853565

[9] Anand Gupta, Harsh Bedi, M.S. Don Bosco, Vinay Shashidhar, "Accuracy of 3D Location Computation in GSM Through NS2," netcom, pp.119-122, 2009 First International Conference on Networks & Communications, 2009.